\documentclass[preprint,showpacs,preprintnumbers,amsmath,amssymb,aps]{revtex4}
\usepackage{graphicx}
\usepackage{dcolumn}
\usepackage{bm}
\usepackage[dvips]{color}
\begin{document}
 
\title{Suppression of hole-mediated ferromagnetism in ${\bf Ga}_{\bf 1-x}{\bf Mn}_{\bf x}{\bf P}$ by hydrogen}
\author{C. Bihler\footnote{bihler@wsi.tum.de}, M. Kraus, and M. S. Brandt}
\affiliation{Walter Schottky Institut, Technische Universit\"{a}t M\"{u}nchen, Am Coulombwall 3, 85748 Garching, Germany}%
\author{S.T.B. Goennenwein and M. Opel}
\affiliation{Walther-Meissner-Institut, Bayerische Akademie der Wissenschaften, Walther-Meissner-Str.~8, 85748 Garching, Germany}
\author{M. A. Scarpulla\footnote{present address: Materials Department, University of California, Santa Barbara, CA 93106, mikes@engineering.ucsb.edu}, R. Farshchi, and O. D. Dubon}
\affiliation{Department of Materials Science and Engineering, University of California, Berkeley and Lawrence Berkeley National Laboratory, Berkeley, CA 94720}

\begin{abstract}
We report the successful passivation of the Mn acceptors in ${\rm Ga}_{1-x}{\rm Mn}_x{\rm P}$ upon exposure to
a remote dc hydrogen plasma. The as-grown films are non-metallic and ferromagnetic
with a Curie temperature of $T_C=55$~K. After hydrogenation the sample resistivity increases
by approximately three orders of magnitude at room temperature and six orders of magnitude at 25 K.
Furthermore, the hydrogenated samples are paramagnetic, which is evidenced by a
magnetization curve at 5 K that is best described by a Brillouin function with $g = 2$ and
$J=5/2$ expected for Mn atoms in the $3d^5$ configuration. These observations unambiguously proof that the ferromagnetism is carrier-mediated also in  ${\rm Ga}_{1-x}{\rm Mn}_x{\rm P}$.
\end{abstract}


\maketitle
The ferromagnetic ordering in Mn-doped diluted magnetic semiconductors (DMS) is generally mediated by holes. In the prototypical III-V DMS ${\rm Ga}_{1-x}{\rm Mn}_x{\rm As}$, the transition metal Mn incorporates substitutionally on the Ga site in the oxidation state 3+ and acts as a relatively shallow acceptor, leading to the formation of Mn$^{2+}$($3d^5$) with a spin $S=5/2$ and a hole $h^+$ in the valence band~\cite{Sch87, Die01}. Therefore, ferromagnetic ${\rm Ga}_{1-x}{\rm Mn}_x{\rm As}$ usually exhibits a metallic conductivity. However, for small ($x\leq0.02$) and high ($x\geq0.07$) Mn concentrations also ferromagnetism on the insulator side of the metal-insulator transition has been reported~\cite{Mat98}. Ferromagnetism without metallic conductivity has also been observed in ${\rm Ga}_{1-x}{\rm Mn}_x{\rm P}$~\cite{Sca05, Dub06, Sto06, Far06, Bih07}. Here, the Mn acceptor level is much deeper in the band gap~\cite{Evw76, Cle85, Gra03a}, so that the concentration of valence band holes is negligible at temperatures below $T_C$. Rather, photoconductivity experiments demonstrate the formation of an acceptor-impurity-band, in which thermally activated hopping processes take place~\cite{Sca05}. This transport is attributed to hopping of highly localized hole states residing in the $t_{2\uparrow}$-orbitals of the Mn $d$-shell, which mediate ferromagnetism.

\begin{figure}
\includegraphics[width=0.7\textwidth]{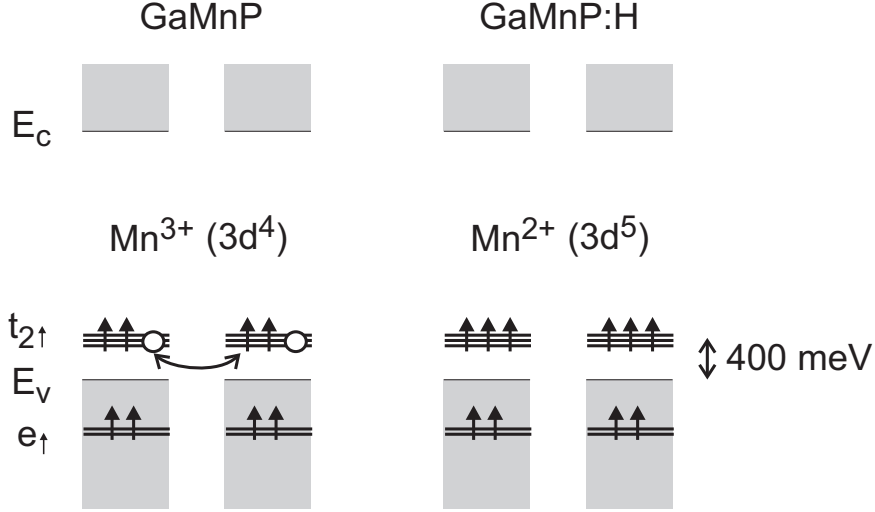}
%\vspace{0.5cm}
\caption{\label{fig:tb} Schematical tight-binding model of the Mn-related energy levels in ${\rm Ga}_{1-x}{\rm Mn}_x{\rm P}$~\cite{Vog85, Gra03a}. The crystal field splits the five $3d$ orbitals into doubly degenerate $e_\uparrow$ and triply degenerate $t_{2\uparrow}$ orbitals. For Mn$^{3+}$, the $t_{2\uparrow}$ orbital is occupied by two electrons only, allowing charge transfer between neighboring $t_{2\uparrow}$ orbitals leading to ferrmagnetic exchange. For Mn$^{2+}$ the $t_{2\uparrow}$ orbitals are full, preventing charge transfer and magnetic ordering.}
\end{figure}

An equivalent description in a pure electron picture is that the magnetic ordering arises from exchange between neighboring Mn$^{3+}$, where transfer of electrons between partially-filled $t_{2\uparrow}$-orbitals is only possible for ferromagnetic ordering due to Hund's rule (Fig.~\ref{fig:tb})~\cite{Vog85, Gra03a}. In this case the ferromagnetic coupling arises from the stabilizing interactions between the partially-occupied orbitals as described by Zhao \textit{et al.}~\cite{Zha05}. If, in contrast, the $t_{2\uparrow}$ orbitals would be fully occupied (corresponding to Mn$^{2+}$), exchange would involve the transient formation of Mn$^{1+}$($3d^6$), which is very unlikely due to the large Hund's coupling energy of typically $>1.5$~eV \cite{Zha00} accounting for the fact that one of the $d$-electrons has a spin which is antiparallel to the remaining $d$-electrons. Therefore, III-V DMS containing exclusively Mn$^{2+}$ are expected to be paramagnetic~\cite{Zha05}. Indeed, the effect of changing the oxidation state of Mn in ${\rm Ga}_{1-x}{\rm Mn}_x{\rm P}$ has already been investigated using partial codoping with donors, where the saturation magnetization and the Curie temperature decrease with increasing compensation~\cite{Sca05, Sto07}. An alternative method to manipulate the oxidation state of Mn was demonstrated for ${\rm Ga}_{1-x}{\rm Mn}_x{\rm As}$ via the incorporation of hydrogen~\cite{Bou03, Goe04,  Bra04}. In this letter we investigate the influence of hydrogenation on the magnetic and electric properties of ${\rm Ga}_{1-x}{\rm Mn}_x{\rm P}$ and show that hydrogenation switches the ferromagnetic state of ${\rm Ga}_{1-x}{\rm Mn}_x{\rm P}$ to a paramagnetic state while further reducing the conductivity. As shown for  ${\rm Ga}_{1-x}{\rm Mn}_x{\rm As}$, such hydrogenation treatments can allow a post-growth tuning of the magnetic properties of DMS thin films including a magnetic patterning while keeping the surface topography unchanged~\cite{The05, Far07, The07b}.

The ${\rm Ga}_{1-x}{\rm Mn}_x{\rm P}$ films studied were prepared by ion implantation followed by pulsed-laser melting (II-PLM) as described elsewhere~\cite{Sca05, Bih07}. This II-PLM processing results in samples having a depth profile of Mn, which$-$as measured by secondary ion mass spectrometry (SIMS)$-$can be approximated by a Gaussian distribution centered at a depth of 40~nm with a width of 20~nm, making it impossible to quote single values for the film thickness and Mn concentration. However, as the regions of the film with highest Mn concentration dominate both the magnetic and transport properties, samples are discussed here in terms of their peak \textit{substitutional} Mn concentration $x=0.042$. Channeling particle induced x-ray emission (PIXE) has shown that about $70\%$ of the total Mn concentration is substitutional~\cite{Yu02, Yu03}. The hydrogenation was carried out via a remote DC hydrogen plasma operated at 0.9 mbar for 30~h, with the sample heated to 170$^\circ$C. This temperature is well below 300$^\circ$C, the temperature above which thermal degradation of the layers has been reported~\cite{Far06}. Since ten samples investigated have shown very similar behavior upon hydrogenation, we focus here on one particular sample already discussed in detail in~\cite{Bih07}, which has a Curie temperature of $T_C=55$~K.

\begin{figure}
\includegraphics[width=0.6\textwidth]{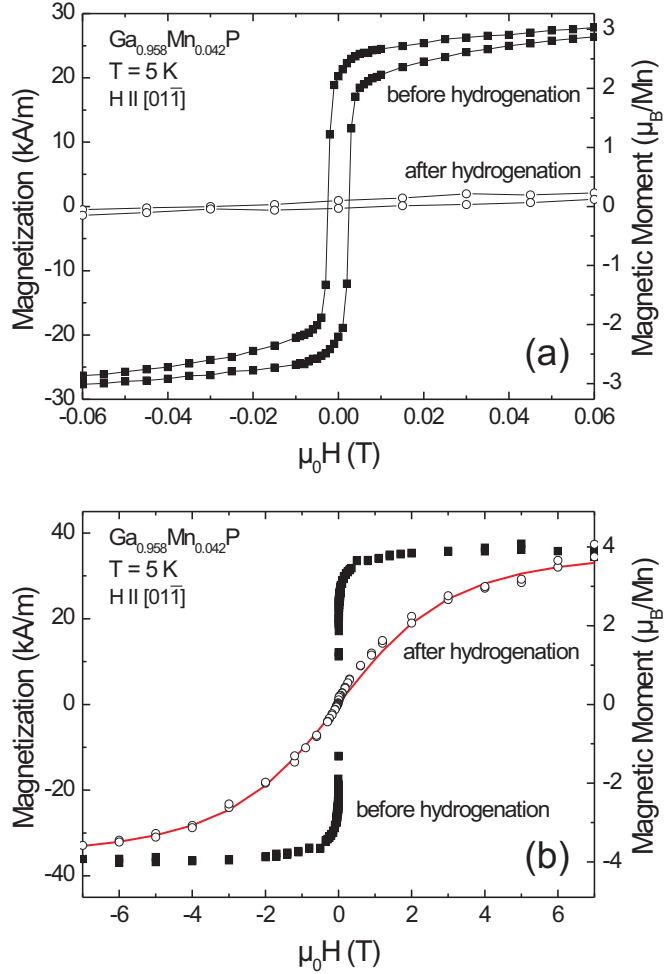}
%\vspace{0.5cm}
\caption{\label{fig:SQUID} (a) Comparison of the dc magnetization $M(H)$ loops of ${\rm Ga}_{0.958}{\rm Mn}_{0.042}{\rm P}$ before (solid squares) and after hydrogenation (open circles) at $T=5$~K with the external magnetic field $H$ applied along the in-plane $\left[01\bar{1}\right]$ axis. The right and left vertical axes give the magnetic moment per \textit{substitutional} Mn atom $m_{\rm Mn}$ and the magnetization $M$ of the sample in the region of highest Mn concentration, respectively. These values were deduced from the measured total magnetic moment $m_{\rm tot}$ as described in Appendix B in Ref.~\cite{Bih07}. The ferromagnetic hysteresis of the ${\rm Ga}_{0.958}{\rm Mn}_{0.042}{\rm P}$ film vanishes upon hydrogenation. (b) Comparison of the same $M(H)$ loops as in (a) on a much broader range of the external magnetic field $H$. The $M(H)$ loop after hydrogenation is well described by a Brillouin-type magnetization with $g=2$ and $J=5/2$ (solid red curve) expected for the paramagnetism of the localized magnetic moments of Mn atoms in the Mn$^{2+}$($3d^5$) state.}
\end{figure}

The drastic changes of the magnetic properties induced by hydrogenation of ${\rm Ga}_{0.958}{\rm Mn}_{0.042}{\rm P}$ are shown in Fig.~\ref{fig:SQUID}. Before hydrogenation the $M(H)$ magnetization loop exhibits a pronounced hysteresis below the Curie temperature $T_C=55$~K. For the external magnetic field $H||\left[01\bar{1}\right]$, which is the magnetic easy axis as shown in~\cite{Bih07}, the coercive field at 5~K is $\mu_0H_C=2.5$~mT (closed squares in Fig.~\ref{fig:SQUID}). The rounded shape of the hysteresis loop, as well as the additional hysteresis observed for $\left| H \right| > H_C$ was explained by the pinning and depinning of domain walls at crystal defects. The magnetization saturates at approximately $4\mu_B$ per substitutional Mn. After hydrogenation ferromagnetism completely vanishes and the $M(H)$ loop (open circles) can be described by a paramagnetic, Brillouin-type magnetization $M(H)=M_{\rm sat}B_J({g\mu_BJ\mu_0H}/{k_BT})$. Here $M_{\rm sat}$ is the saturation magnetization, $B_J$ the Brillouin function, $g$ the $g$ factor, $\mu_B$ the Bohr magneton, $J$ the total angular momentum quantum number, $\mu_0$ the vacuum permeability, and $k_B$ the Boltzmann constant. The saturation magnetization is not significantly changed upon hydrogenation, which shows that the density of localized Mn magnetic moments remains unaffected by the hydrogenation process. The experimental data obtained for the magnetization of the hydrogenated ${\rm Ga}_{1-x}{\rm Mn}_x{\rm P}$ is well described with $g=2$ and $J=5/2$ [full red line in Fig.~\ref{fig:SQUID}(b)]. These parameters can be understood from the oxidation state: Mn with the electron configuration [Ar]$3d^54s^2$ is substituting Ga with [Ar]$3d^{10}4s^24p^1$, where [Ar] is the electron configuration of Ar. The three $4s^24p^1$ valence electrons of Ga contribute to the bonds with the neighboring P atoms, leading to Ga in the oxidation state Ga$^{3+}$ in a fully ionic picture. Since the Mn acceptor level (corresponding to the Mn$^{2+/3+}$ charge transfer level) is too deep in ${\rm Ga}_{1-x}{\rm Mn}_x{\rm P}$, the electron required to satisfy the octet rule for the bonds to the neighboring P atoms cannot be taken out of the valence band. Rather Mn substituting Ga in ${\rm Ga}_{1-x}{\rm Mn}_x{\rm P}$ has to be in the same oxidation state Mn$^{3+}$ $-$ corresponding to the electron configuration [Ar]$3d^4$. Hydrogenation adds a further electron to the Mn which leads to a $3d^5$ configuration with $g\approx2$ and $J=5/2$ \cite{Sch87} corresponding to a Mn$^{2+}$ oxidation state or a fully occupied $t_{2\uparrow}$-orbital.

\begin{figure}
\includegraphics[width=0.7\textwidth]{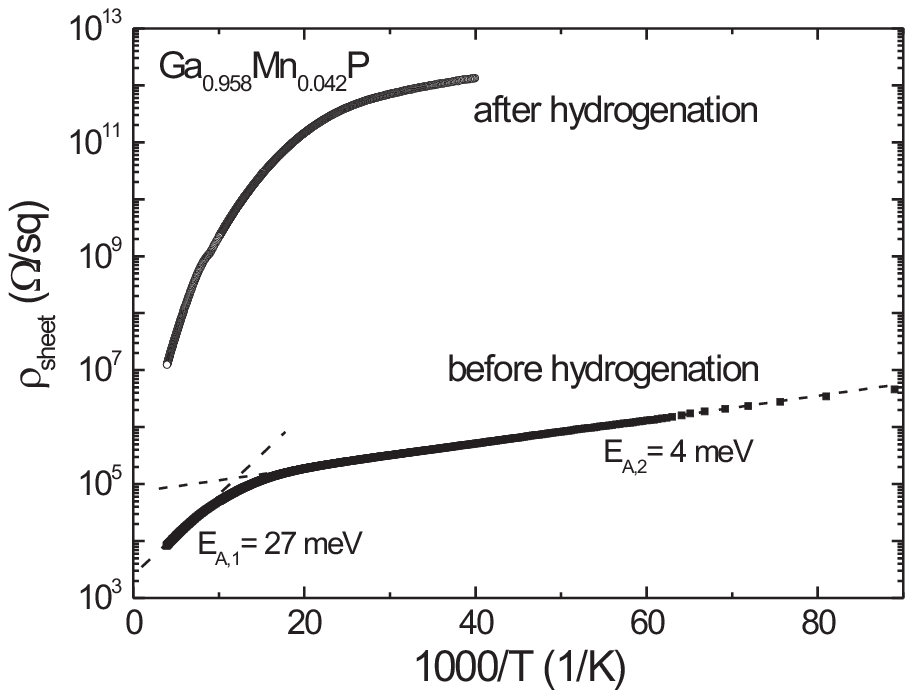}
\vspace{0.5cm}
\caption{\label{fig:resist} Temperature dependence of the dark sheet resistivity before (solid squares) and after hydrogenation (open circles).}
\end{figure}

The change in the oxidation state and the associated full occupancy of the $t_{2\uparrow}$-orbital leads to an electronic passivation of the deep Mn acceptor upon hydrogenation also evidenced by the temperature dependence of dark sheet resistivity $\rho_{\rm sheet}$ in Fig.~\ref{fig:resist}. Before hydrogenation, two thermally activated processes dominate the electronic transport. As described in detail in Ref.~\cite{Sca05}, the high temperature activation energy $E_{\rm A,1}=27$~meV can be attributed to an excitation accross the gap between the Mn-derived impurity band and the valence band, while the smaller low-temperature activation energy $E_{\rm A,2}=4$~meV can be associated with the formation of a continuous hopping transport path in the Mn-impurity band. Upon hydrogenation, $\rho_{\rm sheet}$ increases by three orders of magnitude at room temperature and by six orders of magnitude at 25~K, as expected from the model described above: Via passivation of the holes in the detached impurity band, the resistivity at low temperatures increases strongly. Treatment of ${\rm Ga}_{1-x}{\rm Mn}_x{\rm P}$ with hydrogen has already been reported by Overberg \textit{et al.}~\cite{Ove03}. These authors saw an increase of the Curie temperature upon partial hydrogenation, which they attributed to the passivation of crystal defects and carbon acceptors. Note that they observed a finite magnetization up to 300~K corresponding to a ferromagnetic phase different to the one in our films. Furthermore, the present work shows a switching from ferromagnetism to paramagnetism induced by hydrogenation, which is correlated to a suppression of electronic transport by many orders of magnitude.

In principle, the significant reduction of the conductivity by hydrogenation could be caused by the formation of acceptor-hydrogen complexes or by incorporation of interstitial hydrogen atoms acting as donors and leading to compensation. The position of hydrogen in the crystal is still an open question in ${\rm Ga}_{1-x}{\rm Mn}_x{\rm P}$ as well as in ${\rm Ga}_{1-x}{\rm Mn}_x{\rm As}$. First inferences can be drawn from local vibrational modes. For group-II acceptor-hydrogen complexes in GaAs~\cite{Rah93}, InP~\cite{Dar93}, and GaP~\cite{McC94, McC95} there is an increase in the frequency of the local vibrational modes with increasing atomic number of the group-II acceptor (Be$\rightarrow$Zn$\rightarrow$Cd). This was taken as an indication that in all cases hydrogen binds to the host anion in a $\left\langle 111\right\rangle$ bond-centered orientation~\cite{McC99}, which would explain the influence of the identity of the group-II acceptor on the mode position. Hydrogenated ${\rm Ga}_{1-x}{\rm Mn}_x{\rm As}$ exhibits the same mode positions as hydrogenated Mg- and Zn-doped GaAs~\cite{Bou03,Bra04}. Two independent groups~\cite{Amo05,Gos05,Gia06,Gos06} indeed predict a bond-centered configuration for ${\rm Ga}_{1-x}{\rm Mn}_x{\rm As}{\rm :H}$ to be the energetically favored geometry where the H atom is slightly relaxed away from the Mn-As bond. However, the missing influence of the mode position on the specific acceptor in the case of Mg- and Zn-doped GaAs led Pajot to the alternative suggestion that hydrogen is backbonded to one of the As neighbors in acceptor-hydrogen complexes~\cite{Paj89}. We have not been able to detect the local vibrational modes in the ${\rm Ga}_{1-x}{\rm Mn}_x{\rm P}$ samples studied here by Fourier transform infrared spectroscopy (FTIR), probably due to the high Mn concentration leading to a broadening of the absorption lines. However, the observation of other group-II acceptor-hydrogen complexes in ${\rm Ga}_{1-x}{\rm Mn}_x{\rm P}$ and of Mn-hydrogen in GaAs strongly suggests that irrespective of the exact structure, Mn-hydrogen complexes will be formed, indicating that passivation is the dominant electrical process.

In conclusion we showed that ferromagnetism in ${\rm Ga}_{1-x}{\rm Mn}_x{\rm P}$ can be suppressed via hydrogenation as already reported for ${\rm Ga}_{1-x}{\rm Mn}_x{\rm As}$. In both cases, the effective oxidation state of Mn is changed from 3+ to 2+. In ${\rm Ga}_{1-x}{\rm Mn}_x{\rm As}$ the dominant effect of the full occupation of the $t_{2\uparrow}$-orbitals is the removal of the acceptor level, in   ${\rm Ga}_{1-x}{\rm Mn}_x{\rm P}$ hopping of holes in the acceptor impurity band is suppressed.

Discussions with C. Grasse and P. R. Stone are acknowledged. The work at the Walter Schottky Institut was supported by Deutsche Forschungsgemeinschaft through SFB 631 and the Bavaria California Technology Center, the work at Berkeley by the Director, Office of Science, Office of Basic Energy Sciences, Division of Materials Sciences and Engineering, of the U.S. Department of Energy under Contract No. DEAC02- 05CH11231.


\begin{thebibliography}{34}
\expandafter\ifx\csname natexlab\endcsname\relax\def\natexlab#1{#1}\fi
\expandafter\ifx\csname bibnamefont\endcsname\relax
  \def\bibnamefont#1{#1}\fi
\expandafter\ifx\csname bibfnamefont\endcsname\relax
  \def\bibfnamefont#1{#1}\fi
\expandafter\ifx\csname citenamefont\endcsname\relax
  \def\citenamefont#1{#1}\fi
\expandafter\ifx\csname url\endcsname\relax
  \def\url#1{\texttt{#1}}\fi
\expandafter\ifx\csname urlprefix\endcsname\relax\def\urlprefix{URL }\fi
\providecommand{\bibinfo}[2]{#2}
\providecommand{\eprint}[2][]{\url{#2}}

\bibitem[{\citenamefont{Schneider et~al.}(1987)\citenamefont{Schneider,
  Kaufmann, Wilkening, and Baumler}}]{Sch87}
\bibinfo{author}{\bibfnamefont{J.}~\bibnamefont{Schneider}},
  \bibinfo{author}{\bibfnamefont{U.}~\bibnamefont{Kaufmann}},
  \bibinfo{author}{\bibfnamefont{W.}~\bibnamefont{Wilkening}},
  \bibnamefont{and} \bibinfo{author}{\bibfnamefont{M.}~\bibnamefont{Baumler}},
  \bibinfo{journal}{Phys. Rev. Lett.} \textbf{\bibinfo{volume}{59}},
  \bibinfo{pages}{240} (\bibinfo{year}{1987}).

\bibitem[{\citenamefont{Dietl et~al.}(2001)\citenamefont{Dietl, Ohno, and
  Matsukura}}]{Die01}
\bibinfo{author}{\bibfnamefont{T.}~\bibnamefont{Dietl}},
  \bibinfo{author}{\bibfnamefont{H.}~\bibnamefont{Ohno}}, \bibnamefont{and}
  \bibinfo{author}{\bibfnamefont{F.}~\bibnamefont{Matsukura}},
  \bibinfo{journal}{Phys. Rev. B} \textbf{\bibinfo{volume}{63}},
  \bibinfo{pages}{195205} (\bibinfo{year}{2001}).

\bibitem[{\citenamefont{Matsukura et~al.}(1998)\citenamefont{Matsukura, Ohno,
  Shen, and Sugawara}}]{Mat98}
\bibinfo{author}{\bibfnamefont{F.}~\bibnamefont{Matsukura}},
  \bibinfo{author}{\bibfnamefont{H.}~\bibnamefont{Ohno}},
  \bibinfo{author}{\bibfnamefont{A.}~\bibnamefont{Shen}}, \bibnamefont{and}
  \bibinfo{author}{\bibfnamefont{Y.}~\bibnamefont{Sugawara}},
  \bibinfo{journal}{Phys. Rev. B} \textbf{\bibinfo{volume}{57}},
  \bibinfo{pages}{R2037} (\bibinfo{year}{1998}).

\bibitem[{\citenamefont{Scarpulla et~al.}(2005)\citenamefont{Scarpulla,
  Cardozo, Hlaing~Oo, McCluskey, and Dubon}}]{Sca05}
\bibinfo{author}{\bibfnamefont{M.~A.} \bibnamefont{Scarpulla}},
  \bibinfo{author}{\bibfnamefont{B.~L.} \bibnamefont{Cardozo}},
  \bibinfo{author}{\bibfnamefont{W.~M.} \bibnamefont{Hlaing~Oo}},
  \bibinfo{author}{\bibfnamefont{M.~D.} \bibnamefont{McCluskey}},
  \bibnamefont{and} \bibinfo{author}{\bibfnamefont{O.~D.} \bibnamefont{Dubon}},
  \bibinfo{journal}{Phys. Rev. Lett.} \textbf{\bibinfo{volume}{95}},
  \bibinfo{pages}{207204} (\bibinfo{year}{2005}).

\bibitem[{\citenamefont{Dubon et~al.}(2006)\citenamefont{Dubon, Scarpulla,
  Farshchi, and Yu}}]{Dub06}
\bibinfo{author}{\bibfnamefont{O.~D.} \bibnamefont{Dubon}},
  \bibinfo{author}{\bibfnamefont{M.~A.} \bibnamefont{Scarpulla}},
  \bibinfo{author}{\bibfnamefont{R.}~\bibnamefont{Farshchi}}, \bibnamefont{and}
  \bibinfo{author}{\bibfnamefont{K.~M.} \bibnamefont{Yu}},
  \bibinfo{journal}{Physica B} \textbf{\bibinfo{volume}{376-377}},
  \bibinfo{pages}{630} (\bibinfo{year}{2006}).

\bibitem[{\citenamefont{Stone et~al.}(2006)\citenamefont{Stone, Scarpulla,
  Farshchi, Sharp, Haller, Dubon, Yu, Beeman, Arenholz, Denlinger
  et~al.}}]{Sto06}
\bibinfo{author}{\bibfnamefont{P.~R.} \bibnamefont{Stone}},
  \bibinfo{author}{\bibfnamefont{M.~A.} \bibnamefont{Scarpulla}},
  \bibinfo{author}{\bibfnamefont{R.}~\bibnamefont{Farshchi}},
  \bibinfo{author}{\bibfnamefont{I.~D.} \bibnamefont{Sharp}},
  \bibinfo{author}{\bibfnamefont{E.~E.} \bibnamefont{Haller}},
  \bibinfo{author}{\bibfnamefont{O.~D.} \bibnamefont{Dubon}},
  \bibinfo{author}{\bibfnamefont{K.~M.} \bibnamefont{Yu}},
  \bibinfo{author}{\bibfnamefont{J.~W.} \bibnamefont{Beeman}},
  \bibinfo{author}{\bibfnamefont{E.}~\bibnamefont{Arenholz}},
  \bibinfo{author}{\bibfnamefont{J.~D.} \bibnamefont{Denlinger}},
  \bibnamefont{and} \bibinfo{author}{\bibfnamefont{H.} \bibnamefont{Ohldag}},
  \bibinfo{journal}{Appl. Phys. Lett.}
  \textbf{\bibinfo{volume}{89}}, \bibinfo{pages}{012504}
  (\bibinfo{year}{2006}).

\bibitem[{\citenamefont{Farshchi et~al.}(2006)\citenamefont{Farshchi,
  Scarpulla, Stone, Yu, Sharp, Beeman, Silvestri, Reichertz, Haller, and
  Dubon}}]{Far06}
\bibinfo{author}{\bibfnamefont{R.}~\bibnamefont{Farshchi}},
  \bibinfo{author}{\bibfnamefont{M.~A.} \bibnamefont{Scarpulla}},
  \bibinfo{author}{\bibfnamefont{P.~R.} \bibnamefont{Stone}},
  \bibinfo{author}{\bibfnamefont{K.~M.} \bibnamefont{Yu}},
  \bibinfo{author}{\bibfnamefont{I.~D.} \bibnamefont{Sharp}},
  \bibinfo{author}{\bibfnamefont{J.~W.} \bibnamefont{Beeman}},
  \bibinfo{author}{\bibfnamefont{H.~H.} \bibnamefont{Silvestri}},
  \bibinfo{author}{\bibfnamefont{L.~A.} \bibnamefont{Reichertz}},
  \bibinfo{author}{\bibfnamefont{E.~E.} \bibnamefont{Haller}},
  \bibnamefont{and} \bibinfo{author}{\bibfnamefont{O.~D.} \bibnamefont{Dubon}},
  \bibinfo{journal}{Sol. Stat. Comm.} \textbf{\bibinfo{volume}{140}},
  \bibinfo{pages}{443} (\bibinfo{year}{2006}).

\bibitem[{\citenamefont{Bihler et~al.}(2007)\citenamefont{Bihler, Kraus, Huebl,
  Brandt, Goennenwein, Opel, Scarpulla, Stone, Farshchi, and Dubon}}]{Bih07}
\bibinfo{author}{\bibfnamefont{C.}~\bibnamefont{Bihler}},
  \bibinfo{author}{\bibfnamefont{M.}~\bibnamefont{Kraus}},
  \bibinfo{author}{\bibfnamefont{H.}~\bibnamefont{Huebl}},
  \bibinfo{author}{\bibfnamefont{M.~S.} \bibnamefont{Brandt}},
  \bibinfo{author}{\bibfnamefont{S.~T.~B.} \bibnamefont{Goennenwein}},
  \bibinfo{author}{\bibfnamefont{M.}~\bibnamefont{Opel}},
  \bibinfo{author}{\bibfnamefont{M.~A.} \bibnamefont{Scarpulla}},
  \bibinfo{author}{\bibfnamefont{P.~R.} \bibnamefont{Stone}},
  \bibinfo{author}{\bibfnamefont{R.}~\bibnamefont{Farshchi}}, \bibnamefont{and}
  \bibinfo{author}{\bibfnamefont{O.~D.} \bibnamefont{Dubon}},
  \bibinfo{journal}{Phys. Rev. B} \textbf{\bibinfo{volume}{75}},
  \bibinfo{pages}{214419} (\bibinfo{year}{2007}).

\bibitem[{\citenamefont{Evwaraye and Woodbury}(1976)}]{Evw76}
\bibinfo{author}{\bibfnamefont{A.~O.} \bibnamefont{Evwaraye}} \bibnamefont{and}
  \bibinfo{author}{\bibfnamefont{H.~H.} \bibnamefont{Woodbury}},
  \bibinfo{journal}{J. Appl. Phys.} \textbf{\bibinfo{volume}{47}},
  \bibinfo{pages}{1595} (\bibinfo{year}{1976}).

\bibitem[{\citenamefont{Clerjaud}(1985)}]{Cle85}
\bibinfo{author}{\bibfnamefont{B.}~\bibnamefont{Clerjaud}},
  \bibinfo{journal}{J. Phys. C} \textbf{\bibinfo{volume}{18}},
  \bibinfo{pages}{3615} (\bibinfo{year}{1985}).

\bibitem[{\citenamefont{Graf et~al.}(2003)\citenamefont{Graf, Goennenwein, and
  Brandt}}]{Gra03a}
\bibinfo{author}{\bibfnamefont{T.}~\bibnamefont{Graf}},
  \bibinfo{author}{\bibfnamefont{S.~T.~B.} \bibnamefont{Goennenwein}},
  \bibnamefont{and} \bibinfo{author}{\bibfnamefont{M.~S.}
  \bibnamefont{Brandt}}, \bibinfo{journal}{phys. stat. sol. (b)}
  \textbf{\bibinfo{volume}{239}}, \bibinfo{pages}{277} (\bibinfo{year}{2003}).

\bibitem[{\citenamefont{Vogl}(1985)}]{Vog85}
\bibinfo{author}{\bibfnamefont{P.}~\bibnamefont{Vogl}},
  \bibinfo{journal}{Festk\"{o}rperprobleme} \textbf{\bibinfo{volume}{25}},
  \bibinfo{pages}{563} (\bibinfo{year}{1985}).

\bibitem[{\citenamefont{Zhao et~al.}(2005)\citenamefont{Zhao, Mahadevan, and
  Zunger}}]{Zha05}
\bibinfo{author}{\bibfnamefont{Y.}~\bibnamefont{Zhao}},
  \bibinfo{author}{\bibfnamefont{P.}~\bibnamefont{Mahadevan}},
  \bibnamefont{and} \bibinfo{author}{\bibfnamefont{A.}~\bibnamefont{Zunger}},
  \bibinfo{journal}{J. Appl. Phys.} \textbf{\bibinfo{volume}{98}},
  \bibinfo{pages}{113901} (\bibinfo{year}{2005}).

\bibitem[{\citenamefont{Zhao}(2000)}]{Zha00}
\bibinfo{author}{\bibfnamefont{G.}~\bibnamefont{Zhao}}, \bibinfo{journal}{Phys.
  Rev. B} \textbf{\bibinfo{volume}{62}}, \bibinfo{pages}{11639}
  (\bibinfo{year}{2000}).

\bibitem[{\citenamefont{Stone et~al.}(2007)\citenamefont{Stone, Scarpulla,
  Farshchi, Sharp, Beeman, Yu, Arenholz, Denlinger, Haller, and Dubon}}]{Sto07}
\bibinfo{author}{\bibfnamefont{P.~R.} \bibnamefont{Stone}},
  \bibinfo{author}{\bibfnamefont{M.~A.} \bibnamefont{Scarpulla}},
  \bibinfo{author}{\bibfnamefont{R.}~\bibnamefont{Farshchi}},
  \bibinfo{author}{\bibfnamefont{I.~D.} \bibnamefont{Sharp}},
  \bibinfo{author}{\bibfnamefont{J.~W.} \bibnamefont{Beeman}},
  \bibinfo{author}{\bibfnamefont{K.~M.} \bibnamefont{Yu}},
  \bibinfo{author}{\bibfnamefont{E.}~\bibnamefont{Arenholz}},
  \bibinfo{author}{\bibfnamefont{J.~D.} \bibnamefont{Denlinger}},
  \bibinfo{author}{\bibfnamefont{E.~E.} \bibnamefont{Haller}},
  \bibnamefont{and} \bibinfo{author}{\bibfnamefont{O.~D.} \bibnamefont{Dubon}},
  \bibinfo{journal}{AIP Conf. Proc.} \textbf{\bibinfo{volume}{893}},
  \bibinfo{pages}{1177} (\bibinfo{year}{2007}).

\bibitem[{\citenamefont{Bouanani-Rahbi
  et~al.}(2003)\citenamefont{Bouanani-Rahbi, Clerjaud, Theys, Lema$\hat{\rm
  i}$tre, and Jomard}}]{Bou03}
\bibinfo{author}{\bibfnamefont{R.}~\bibnamefont{Bouanani-Rahbi}},
  \bibinfo{author}{\bibfnamefont{B.}~\bibnamefont{Clerjaud}},
  \bibinfo{author}{\bibfnamefont{B.}~\bibnamefont{Theys}},
  \bibinfo{author}{\bibfnamefont{A.}~\bibnamefont{Lema$\hat{\rm i}$tre}},
  \bibnamefont{and} \bibinfo{author}{\bibfnamefont{F.}~\bibnamefont{Jomard}},
  \bibinfo{journal}{Physica B} \textbf{\bibinfo{volume}{340-342}},
  \bibinfo{pages}{284} (\bibinfo{year}{2003}).

\bibitem[{\citenamefont{Goennenwein et~al.}(2004)\citenamefont{Goennenwein,
  Wassner, Huebl, Brandt, Philipp, Opel, Gross, Koeder, Schoch, and
  Waag}}]{Goe04}
\bibinfo{author}{\bibfnamefont{S.~T.~B.} \bibnamefont{Goennenwein}},
  \bibinfo{author}{\bibfnamefont{T.~A.} \bibnamefont{Wassner}},
  \bibinfo{author}{\bibfnamefont{H.}~\bibnamefont{Huebl}},
  \bibinfo{author}{\bibfnamefont{M.~S.} \bibnamefont{Brandt}},
  \bibinfo{author}{\bibfnamefont{J.~B.} \bibnamefont{Philipp}},
  \bibinfo{author}{\bibfnamefont{M.}~\bibnamefont{Opel}},
  \bibinfo{author}{\bibfnamefont{R.}~\bibnamefont{Gross}},
  \bibinfo{author}{\bibfnamefont{A.}~\bibnamefont{Koeder}},
  \bibinfo{author}{\bibfnamefont{W.}~\bibnamefont{Schoch}}, \bibnamefont{and}
  \bibinfo{author}{\bibfnamefont{A.}~\bibnamefont{Waag}},
  \bibinfo{journal}{Phys. Rev. Lett.} \textbf{\bibinfo{volume}{92}},
  \bibinfo{pages}{227202} (\bibinfo{year}{2004}).

\bibitem[{\citenamefont{Brandt et~al.}(2004)\citenamefont{Brandt, Goennenwein,
  Wassner, Kohl, Lehner, Huebl, Graf, Stutzmann, Koeder, Schoch
  et~al.}}]{Bra04}
\bibinfo{author}{\bibfnamefont{M.~S.} \bibnamefont{Brandt}},
  \bibinfo{author}{\bibfnamefont{S.~T.~B.} \bibnamefont{Goennenwein}},
  \bibinfo{author}{\bibfnamefont{T.~A.} \bibnamefont{Wassner}},
  \bibinfo{author}{\bibfnamefont{F.}~\bibnamefont{Kohl}},
  \bibinfo{author}{\bibfnamefont{A.}~\bibnamefont{Lehner}},
  \bibinfo{author}{\bibfnamefont{H.}~\bibnamefont{Huebl}},
  \bibinfo{author}{\bibfnamefont{T.}~\bibnamefont{Graf}},
  \bibinfo{author}{\bibfnamefont{M.}~\bibnamefont{Stutzmann}},
  \bibinfo{author}{\bibfnamefont{A.}~\bibnamefont{Koeder}},
  \bibinfo{author}{\bibfnamefont{W.}~\bibnamefont{Schoch}},
  \bibnamefont{and} \bibinfo{author}{\bibfnamefont{A.}~\bibnamefont{Waag}},
  \bibinfo{journal}{Appl. Phys. Lett.}
  \textbf{\bibinfo{volume}{84}}, \bibinfo{pages}{2277} (\bibinfo{year}{2004}).

\bibitem[{\citenamefont{Thevenard et~al.}(2005)\citenamefont{Thevenard,
  Largeau, Mauguin, Lema$\hat{\rm i}$tre, and Theys}}]{The05}
\bibinfo{author}{\bibfnamefont{L.}~\bibnamefont{Thevenard}},
  \bibinfo{author}{\bibfnamefont{L.}~\bibnamefont{Largeau}},
  \bibinfo{author}{\bibfnamefont{O.}~\bibnamefont{Mauguin}},
  \bibinfo{author}{\bibfnamefont{A.}~\bibnamefont{Lema$\hat{\rm i}$tre}},
  \bibnamefont{and} \bibinfo{author}{\bibfnamefont{B.}~\bibnamefont{Theys}},
  \bibinfo{journal}{Appl. Phys. Lett.} \textbf{\bibinfo{volume}{87}},
  \bibinfo{pages}{182506} (\bibinfo{year}{2005}).

\bibitem[{\citenamefont{Farshchi et~al.}(2007)\citenamefont{Farshchi,
  Chopdekar, Suzuki, Ashby, Sharp, Beeman, Haller, and Dubon}}]{Far07}
\bibinfo{author}{\bibfnamefont{R.}~\bibnamefont{Farshchi}},
  \bibinfo{author}{\bibfnamefont{R.~V.} \bibnamefont{Chopdekar}},
  \bibinfo{author}{\bibfnamefont{Y.}~\bibnamefont{Suzuki}},
  \bibinfo{author}{\bibfnamefont{P.~D.} \bibnamefont{Ashby}},
  \bibinfo{author}{\bibfnamefont{I.~D.} \bibnamefont{Sharp}},
  \bibinfo{author}{\bibfnamefont{J.~W.} \bibnamefont{Beeman}},
  \bibinfo{author}{\bibfnamefont{E.~E.} \bibnamefont{Haller}},
  \bibnamefont{and} \bibinfo{author}{\bibfnamefont{O.~D.} \bibnamefont{Dubon}},
  \bibinfo{journal}{phys. stat. sol. (c)} \textbf{\bibinfo{volume}{4}},
  \bibinfo{pages}{1755} (\bibinfo{year}{2007}).

\bibitem[{\citenamefont{Thevenard et~al.}(2007)\citenamefont{Thevenard, Miard,
  Vila, Faini, Lema$\hat{\rm i}$tre, Vernier, Ferr$\acute{\rm e}$, and
  Fusil}}]{The07b}
\bibinfo{author}{\bibfnamefont{L.}~\bibnamefont{Thevenard}},
  \bibinfo{author}{\bibfnamefont{A.}~\bibnamefont{Miard}},
  \bibinfo{author}{\bibfnamefont{L.}~\bibnamefont{Vila}},
  \bibinfo{author}{\bibfnamefont{G.}~\bibnamefont{Faini}},
  \bibinfo{author}{\bibfnamefont{A.}~\bibnamefont{Lema$\hat{\rm i}$tre}},
  \bibinfo{author}{\bibfnamefont{N.}~\bibnamefont{Vernier}},
  \bibinfo{author}{\bibfnamefont{J.}~\bibnamefont{Ferr$\acute{\rm e}$}},
  \bibnamefont{and} \bibinfo{author}{\bibfnamefont{S.}~\bibnamefont{Fusil}},
  \bibinfo{journal}{arXiv:0706.2138v1}  (\bibinfo{year}{2007}).

\bibitem[{\citenamefont{Yu et~al.}(2002)\citenamefont{Yu, Walukiewicz,
  Wojtowicz, Kuryliszyn, Liu, Sasaki, and Furdyna}}]{Yu02}
\bibinfo{author}{\bibfnamefont{K.~M.} \bibnamefont{Yu}},
  \bibinfo{author}{\bibfnamefont{W.}~\bibnamefont{Walukiewicz}},
  \bibinfo{author}{\bibfnamefont{T.}~\bibnamefont{Wojtowicz}},
  \bibinfo{author}{\bibfnamefont{I.}~\bibnamefont{Kuryliszyn}},
  \bibinfo{author}{\bibfnamefont{X.}~\bibnamefont{Liu}},
  \bibinfo{author}{\bibfnamefont{Y.}~\bibnamefont{Sasaki}}, \bibnamefont{and}
  \bibinfo{author}{\bibfnamefont{J.~K.} \bibnamefont{Furdyna}},
  \bibinfo{journal}{Phys. Rev. B} \textbf{\bibinfo{volume}{65}},
  \bibinfo{pages}{201303} (\bibinfo{year}{2002}).

\bibitem[{\citenamefont{Yu et~al.}(Yu022003)\citenamefont{Yu, Walukiewicz,
  Wojtowicz, Lim, Liu, Bindley, Dobrowolska, and Furdyna}}]{Yu03}
\bibinfo{author}{\bibfnamefont{K.~M.} \bibnamefont{Yu}},
  \bibinfo{author}{\bibfnamefont{W.}~\bibnamefont{Walukiewicz}},
  \bibinfo{author}{\bibfnamefont{T.}~\bibnamefont{Wojtowicz}},
  \bibinfo{author}{\bibfnamefont{W.~L.} \bibnamefont{Lim}},
  \bibinfo{author}{\bibfnamefont{X.}~\bibnamefont{Liu}},
  \bibinfo{author}{\bibfnamefont{U.}~\bibnamefont{Bindley}},
  \bibinfo{author}{\bibfnamefont{M.}~\bibnamefont{Dobrowolska}},
  \bibnamefont{and} \bibinfo{author}{\bibfnamefont{J.~K.}
  \bibnamefont{Furdyna}}, \bibinfo{journal}{Phys. Rev. B}
  \textbf{\bibinfo{volume}{68}}, \bibinfo{pages}{041308}
  (\bibinfo{year}{Yu022003}).

\bibitem[{\citenamefont{Overberg et~al.}(2003)\citenamefont{Overberg, Baik,
  Thaler, Abernathy, Pearton, Kelly, Rairigh, Hebard, Tang, Stavola
  et~al.}}]{Ove03}
\bibinfo{author}{\bibfnamefont{M.~E.} \bibnamefont{Overberg}},
  \bibinfo{author}{\bibfnamefont{K.~H.} \bibnamefont{Baik}},
  \bibinfo{author}{\bibfnamefont{G.~T.} \bibnamefont{Thaler}},
  \bibinfo{author}{\bibfnamefont{C.~R.} \bibnamefont{Abernathy}},
  \bibinfo{author}{\bibfnamefont{S.~J.} \bibnamefont{Pearton}},
  \bibinfo{author}{\bibfnamefont{J.}~\bibnamefont{Kelly}},
  \bibinfo{author}{\bibfnamefont{R.}~\bibnamefont{Rairigh}},
  \bibinfo{author}{\bibfnamefont{A.~F.} \bibnamefont{Hebard}},
  \bibinfo{author}{\bibfnamefont{W.}~\bibnamefont{Tang}},
  \bibinfo{author}{\bibfnamefont{M.}~\bibnamefont{Stavola}},
  \bibnamefont{and} \bibinfo{author}{\bibfnamefont{J.~M.}~\bibnamefont{Zavada}},
  \bibinfo{journal}{Electrochem. Sol.-Stat. Lett.}
  \textbf{\bibinfo{volume}{6}}, \bibinfo{pages}{131} (\bibinfo{year}{2003}).

\bibitem[{\citenamefont{Rahbi et~al.}(1993)\citenamefont{Rahbi, Pajot,
  Chevallier, Marbeuf, Logan, and Gavand}}]{Rah93}
\bibinfo{author}{\bibfnamefont{R.}~\bibnamefont{Rahbi}},
  \bibinfo{author}{\bibfnamefont{B.}~\bibnamefont{Pajot}},
  \bibinfo{author}{\bibfnamefont{J.}~\bibnamefont{Chevallier}},
  \bibinfo{author}{\bibfnamefont{A.}~\bibnamefont{Marbeuf}},
  \bibinfo{author}{\bibfnamefont{R.~C.} \bibnamefont{Logan}}, \bibnamefont{and}
  \bibinfo{author}{\bibfnamefont{M.}~\bibnamefont{Gavand}},
  \bibinfo{journal}{J. Appl. Phys.} \textbf{\bibinfo{volume}{73}},
  \bibinfo{pages}{1723} (\bibinfo{year}{1993}).

\bibitem[{\citenamefont{Darwich et~al.}(1993)\citenamefont{Darwich, Pajot,
  Rose, Robein, Theys, Rahbi, Porte, and Gendron}}]{Dar93}
\bibinfo{author}{\bibfnamefont{R.}~\bibnamefont{Darwich}},
  \bibinfo{author}{\bibfnamefont{B.}~\bibnamefont{Pajot}},
  \bibinfo{author}{\bibfnamefont{B.}~\bibnamefont{Rose}},
  \bibinfo{author}{\bibfnamefont{D.}~\bibnamefont{Robein}},
  \bibinfo{author}{\bibfnamefont{B.}~\bibnamefont{Theys}},
  \bibinfo{author}{\bibfnamefont{R.}~\bibnamefont{Rahbi}},
  \bibinfo{author}{\bibfnamefont{C.}~\bibnamefont{Porte}}, \bibnamefont{and}
  \bibinfo{author}{\bibfnamefont{F.}~\bibnamefont{Gendron}},
  \bibinfo{journal}{Phys. Rev. B} \textbf{\bibinfo{volume}{48}},
  \bibinfo{pages}{17776} (\bibinfo{year}{1993}).

\bibitem[{\citenamefont{McCluskey et~al.}(1994)\citenamefont{McCluskey, Haller,
  Walker, and Johnson}}]{McC94}
\bibinfo{author}{\bibfnamefont{M.~D.} \bibnamefont{McCluskey}},
  \bibinfo{author}{\bibfnamefont{E.~E.} \bibnamefont{Haller}},
  \bibinfo{author}{\bibfnamefont{J.}~\bibnamefont{Walker}}, \bibnamefont{and}
  \bibinfo{author}{\bibfnamefont{N.~M.} \bibnamefont{Johnson}},
  \bibinfo{journal}{Appl. Phys. Lett.} \textbf{\bibinfo{volume}{65}},
  \bibinfo{pages}{2191} (\bibinfo{year}{1994}).

\bibitem[{\citenamefont{McCluskey et~al.}(1995)\citenamefont{McCluskey, Haller,
  Walker, and Johnson}}]{McC95}
\bibinfo{author}{\bibfnamefont{M.~D.} \bibnamefont{McCluskey}},
  \bibinfo{author}{\bibfnamefont{E.~E.} \bibnamefont{Haller}},
  \bibinfo{author}{\bibfnamefont{J.}~\bibnamefont{Walker}}, \bibnamefont{and}
  \bibinfo{author}{\bibfnamefont{N.~M.} \bibnamefont{Johnson}},
  \bibinfo{journal}{Phys. Rev. B} \textbf{\bibinfo{volume}{52}},
  \bibinfo{pages}{11859} (\bibinfo{year}{1995}).

\bibitem[{\citenamefont{McCluskey and Haller}(1999)}]{McC99}
\bibinfo{author}{\bibfnamefont{M.~D.} \bibnamefont{McCluskey}}
  \bibnamefont{and} \bibinfo{author}{\bibfnamefont{E.~E.}
  \bibnamefont{Haller}}, \emph{\bibinfo{title}{Hydrogen in Semiconductors II}},
  vol. \bibinfo{volume}{61, chap. 9, ed. by N. H. Nickel}
  (\bibinfo{publisher}{Academic Press, San Diego}, \bibinfo{year}{1999}).

\bibitem[{\citenamefont{Amore~Bonapasta
  et~al.}(2005)\citenamefont{Amore~Bonapasta, Filippone, and
  Giannozzi}}]{Amo05}
\bibinfo{author}{\bibfnamefont{A.}~\bibnamefont{Amore~Bonapasta}},
  \bibinfo{author}{\bibfnamefont{F.}~\bibnamefont{Filippone}},
  \bibnamefont{and}
  \bibinfo{author}{\bibfnamefont{P.}~\bibnamefont{Giannozzi}},
  \bibinfo{journal}{Phys. Rev. B} \textbf{\bibinfo{volume}{72}},
  \bibinfo{pages}{121202} (\bibinfo{year}{2005}).

\bibitem[{\citenamefont{Goss and Briddon}(2005)}]{Gos05}
\bibinfo{author}{\bibfnamefont{J.~P.} \bibnamefont{Goss}} \bibnamefont{and}
  \bibinfo{author}{\bibfnamefont{P.~R.} \bibnamefont{Briddon}},
  \bibinfo{journal}{Phys. Rev. B} \textbf{\bibinfo{volume}{72}},
  \bibinfo{pages}{115211} (\bibinfo{year}{2005}).

\bibitem[{\citenamefont{Giannozzi et~al.}(2006)\citenamefont{Giannozzi,
  Filippone, and Amore~Bonapasta}}]{Gia06}
\bibinfo{author}{\bibfnamefont{P.}~\bibnamefont{Giannozzi}},
  \bibinfo{author}{\bibfnamefont{F.}~\bibnamefont{Filippone}},
  \bibnamefont{and}
  \bibinfo{author}{\bibfnamefont{A.}~\bibnamefont{Amore~Bonapasta}},
  \bibinfo{journal}{Braz. J. Phys.} \textbf{\bibinfo{volume}{36}},
  \bibinfo{pages}{245} (\bibinfo{year}{2006}).

\bibitem[{\citenamefont{Goss et~al.}(2006)\citenamefont{Goss, Briddon, and
  Wardle}}]{Gos06}
\bibinfo{author}{\bibfnamefont{J.~P.} \bibnamefont{Goss}},
  \bibinfo{author}{\bibfnamefont{P.~R.} \bibnamefont{Briddon}},
  \bibnamefont{and} \bibinfo{author}{\bibfnamefont{M.~G.}
  \bibnamefont{Wardle}}, \bibinfo{journal}{Physica B}
  \textbf{\bibinfo{volume}{376-377}}, \bibinfo{pages}{654}
  (\bibinfo{year}{2006}).

\bibitem[{\citenamefont{Pajot}(1989)}]{Paj89}
\bibinfo{author}{\bibfnamefont{B.}~\bibnamefont{Pajot}},
  \bibinfo{journal}{Inst. Phys. Conf. Ser.} \textbf{\bibinfo{volume}{95}},
  \bibinfo{pages}{437} (\bibinfo{year}{1989}).

\end{thebibliography}
\end{document}